\documentclass[12pt]{article}
\usepackage[utf8]{inputenc}
\usepackage{cite}
\usepackage{graphicx,color}
\usepackage{setspace}
\usepackage{nicefrac}
\usepackage[nottoc]{tocbibind} 

\usepackage[linktocpage=true,colorlinks=true,linkcolor=blue,citecolor=blue,urlcolor=blue]{hyperref}

\usepackage{newtxtext,newtxmath}

\DeclareMathAlphabet{\mathcal}{OMS}{cmsy}{m}{n}

\usepackage{geometry}
\newcommand{\m}{\mu}
\newcommand{\n}{\nu}

\newcommand{\el}{\ell}
\newcommand{\ep}{\epsilon}

\newcommand{\pr}{\prime}

\newcommand{\s}{\sigma}

\newcommand{\z}{\zeta}
\newcommand{\nn}{\nonumber}

\newcommand{\goesto}{\rightarrow}
\newcommand{\cR}{\mathcal{R}}
\newcommand{\cL}{\mathcal{L}}
\newcommand{\cH}{\mathcal{H}}

\newcommand{\der}{\nabla}
\newcommand{\dif}{\mathrm{d}}
\newcommand{\link}[1]{[\href{http://arxiv.org/abs/#1}{{\tt arXiv:#1}}]}
\newcommand{\linkth}[1]{[\href{http://arxiv.org/abs/hep-th/#1}{{\tt arXiv/hep-th:#1}}]}
\newcommand{\linkgr}[1]{[\href{http://arxiv.org/abs/gr-qc/#1}{{\tt arXiv/gr-qc:#1}}]}
\newcommand{\doi}[1]{[\href{http://doi.org/#1}{{\tt doi:#1}}]}

\numberwithin{equation}{section}

\usepackage{titletoc}

\begin{document}

\begin{titlepage}   
	 

\begin{center}
	\setstretch{2}
	{\LARGE \bf Lower-dimensional Limits of \\Cubic Lovelock Gravity}
\end{center}

\begin{center}

{Gökhan Alkaç${}^a$, Gökçen Deniz Özen${}^b$, Gün Süer${}^c$}\\
{\singlespacing
{\small 
{\it ${}^a$Physics Engineering Department, Faculty of Engineering,\\ Hacettepe University, 06800, Ankara, Turkey}\\[2mm]

{\it ${}^b$Department of Physics, Faculty of Science and Letters,\\ Istanbul Technical University, Maslak 34469 Istanbul, Turkey}\\[2mm]

{\it ${}^c$Department of Physics, Faculty of Arts and Sciences,\\ Middle East Technical University, 06800, Ankara, Turkey}\\[2mm]

e-mail: {\tt{gkhnalkac@gmail.com}, \tt{gd.ozen@gmail.com}, \tt{suer.gun@metu.edu.tr}}
}
}
\end{center}
\begin{center}
{\bf \large Abstract} 
\end{center}

\noindent We obtain the lower-dimensional limits $(p=2,3,4,5,6)$ of cubic Lovelock gravity through a regularized Kaluza-Klein reduction. By taking a flat internal space for simplicity, we also study the static black hole solutions in the resulting theories. We show that the solutions match with the ones obtained from the ``naive limit'' of $D$-dimensional equation for the metric function, which is obtained by first scaling the relevant couplings by a factor of $\frac{1}{D-p}$ and then taking the limit $D\to p$, with one important exception: In 4D, one obtains the expected solution only for the black hole with a planar horizon.

\newpage
\thispagestyle{empty}
\tableofcontents
\end{titlepage}

\section{Introduction}
Presumably, the most straightforward way to find a generalization of Einstein's equations is to start from an action in the following form
\begin{equation}
\label{eqn1}
S=\int \dif^D x\,\sqrt{-g} \,\mathcal{L}\,[g^{\mu\nu}, R^{\mu}\,_{\nu\rho\sigma}],
\end{equation}
whose variation yields the field equations
\begin{align}
\varepsilon_{\mu\nu}
&=\frac{1}{\sqrt{-g}}\frac{\partial(\sqrt{-g}\mathcal{L})}{\partial g^{\mu\nu}}-2\nabla^{\alpha}\nabla^{\beta}P_{\mu\alpha\beta \nu},\label{eq:1.2}\\
&=\cR_{\mu\nu}-\frac{1}{2}g_{\mu\nu}\mathcal{L}-2\nabla^{\alpha}\nabla^{\beta}P_{\mu\alpha\beta\nu}=0,\label{eq:1.3}
\end{align}
where $P_{\mu\nu\rho\sigma}=\dfrac{\partial \mathcal{L}}
{\partial R^{\mu\nu\rho\sigma}}$ and $\cR_{\mu\nu}=P_\mu\,^{\alpha\beta\gamma}R_{\nu\alpha\beta\gamma}$\cite{Padmanabhan:2011ex}. Derived from an action, the tensor $\varepsilon_{\mu\nu}$ is symmetric and divergence-free. Therefore; it can be coupled to an energy-momentum tensor as $\varepsilon_{\mu\nu} \propto T_{\mu\nu}$, yielding a generalization of Einstein's equations. However, the last term in \eqref{eq:1.3} contains derivatives of the metric tensor of order higher than two, which in general spoils the unitarity of the theory. Demanding second-order field equations $(\nabla_\alpha P^{\alpha\mu\nu\beta}=0)$ gives a very special class of theories \cite{Padmanabhan:2013xyr}, Lovelock gravity\footnote{see Section 15.4 of \cite{Padmanabhan:2010zzb} for a pedagogical exposition.} \cite{Lanczos:1938sf,Lovelock:1971yv,Lovelock:1972vz}, that possesses a unitary massless spin-$2$ excitation around any of its constant curvature vacua \cite{Sisman:2012rc}. The Lagrangian of Lovelock gravity is given by
\begin{equation}
\label{eqn3}
\mathcal{L}=\sum_{m} c_m \mathcal{L}_m,
\end{equation}
where the $m$-th order Lovelock Lagrangian reads
\begin{equation}
\label{eqn4}
\mathcal{L}_m=\frac{1}{2^m}\delta^{\mu_1\nu_1\cdots\mu_m\nu_m}_{\rho_1\sigma_1\cdots\rho_m\sigma_m}R^{\rho_1\sigma_1}_{\ \ \ \ \mu_1\nu_1}\cdots R^{\rho_m\sigma_m}_{\ \ \ \ \mu_m\nu_m}.
\end{equation}
Due to the anti-symmetrization in the generalized Kronecker delta symbol, $\mathcal{L}_m$'s identically vanish in $ D\le 2m-1$ and $D=2m$ is the critical dimension where $\mathcal{L}_m$ is a boundary term. As a result, a non-trivial contribution to field equations is possible only when $ D \ge 2m+1$. Unfortunately, in $D=4$, this leaves $ \mathcal{L}_{m=1}=R$, the Einstein-Hilbert term as the only possibility, providing no generalization.

Recently, it was claimed in \cite{Glavan:2019inb} that this can be circumvented for Einstein-Gauss-Bonnet (EGB) theory defined by the Lagrangian
\begin{equation}
\label{eqn5}
\mathcal{L}=\mathcal{L}_{m=1} +\alpha L^2\mathcal{L}_{m=2}=R+\alpha L^2\left(R^2-4R_{\mu\nu}^2+R_{\mu\nu\rho\sigma}^2\right),
\end{equation}
as follows: One can start from a $D$-dimensional theory and a $D$-dimensional metric ansatz for a solution of desired type. The contribution to the field equations from the Gauss-Bonnet term, $\mathcal{L}_{m=2}$, carries a factor of $D-4$. Scaling the coupling as $\alpha \to \frac{\alpha}{D-4}$ and then taking the limit $D\rightarrow 4$ yields a novel 4D solution. In \cite{Glavan:2019inb}, the authors studied constant curvature spacetimes, the cosmological spacetimes, the spherically symmetric black holes and the linearized fluctuations around maximally symmetric vacua. However, the Lagrangian now becomes 
\begin{equation}
\label{eqn6}
\mathcal{L}=\mathcal{L}_{m=1} +\frac{\alpha L^2}{D-4} \mathcal{L}_{m=2},
\end{equation}
which diverges as $D\rightarrow 4$. An immediate problem arising from this can be seen as follows \cite{Lu:2020iav}: For the spherically symmetric black hole, the entropy computed from Iyer-Wald formula \cite{Wald:1993nt,Iyer:1994ys}
\begin{equation}
\label{eqn7}
S\propto \int_{\cH} \dif^{D-2} \sqrt{|h|} P^{\mu\nu\rho\sigma}\epsilon_{\mu\nu}\epsilon_{\rho\sigma},
\end{equation}
where $h$ is the determinant of the induced metric on the horizon and $\epsilon_{\mu\nu}$ is the binormal to the event horizon, diverges as $D\rightarrow 4$. Many other inconsistencies of this ``naive limit'' have been presented in the literature and it is now certain that it does not yield a consistent generalization of the Einstein's theory \cite{Gurses:2020ofy,Gurses:2020rxb,Arrechea:2020evj,Arrechea:2020gjw,Ai:2020peo,Fernandes:2020nbq,Mahapatra:2020rds,Bonifacio:2020vbk,Aoki:2020lig,Shu:2020cjw,Hohmann:2020cor,Cao:2021nng}.

Later it was realized that a rigorous limit to lower-dimensions ($D \goesto p=2,3,4$ for EGB theory) can still be defined; however, the resulting theory is, instead of a pure gravity theory, a scalar-tensor theory \cite{Fernandes:2020nbq,Lu:2020iav,Kobayashi:2020wqy,Hennigar:2020lsl}. Having field equations with at most second derivatives, it is an example of Horndeski gravity \cite{Horndeski:1974wa,Kobayashi:2019hrl} or generalized Galileons \cite{Deffayet:2009mn,Charmousis:2012dw}. Various aspects of these theories have been studied in \cite{Ma:2020ufk,Hennigar:2020drx,Hennigar:2020fkv} (see \cite{Fernandes:2022zrq} for an extensive review of the literature).

Although the novel lower-dimensional solutions obtained from the naive limit are interesting for gravitational physics in the context of black holes, cosmology and weak-field gravity, it is absolutely crucial to obtain them as solutions of well-defined scalar-tensor theories, because otherwise it is impossible to give them a proper physical meaning. For example, a finite entropy for the static black hole solution can only be obtained this way (see \cite{Lu:2020iav} for $D=4$ and \cite{Hennigar:2020drx,Hennigar:2020fkv} for $D=3$).

While the above-mentioned solutions of physical interest can be realized in a well-defined scalar tensor theory in the case of EGB theory, any result beyond the Gauss-Bonnet term is missing in the literature. In \cite{Konoplya:2020ibi}, the static black hole solution of Lovelock gravity of \emph{arbitrary} order is given in $D=3$ by employing the naive limit; however, it is by no means guaranteed that a scalar-tensor theory admitting this solution exists. In order to better understand the nature and the validity of the solutions obtained by the naive limit, we will perform the most obvious nontrivial check and obtain the lower-dimensional limits ($D \goesto p=2, 3, 4, 5, 6$) of cubic Lovelock gravity via a regularized Kaluza-Klein (KK) reduction, extending the analysis of \cite{Lu:2020iav,Kobayashi:2020wqy} for EGB theory. We will check whether the static solution obtained from the naive limit survives in these scalar-tensor theories, and show that
all the static black hole solutions with different horizon topologies (spherical, flat, hyperbolic) are preserved\footnote{The solution presented in \cite{Konoplya:2020ibi} is a solution of the theory we present in subsection \ref{subsec:3d}.} with an important exception: In 4D, only a planar horizon is allowed. This is, to our knowledge, the only example where some solutions are excluded when required to be admitted by a well-defined scalar-tensor theory.

The outline of this paper is as follows: After a discussion of the naive limit and the KK procedure in Section \ref{sec:2}, we present the resulting lower-dimensional theories\footnote{During the completion of this paper, \cite{Matsumoto:2022fln} appeared where the authors considered the regularized KK reduction of Lovelock gravity with flat internal space by analytically continuing the previous results of \cite{VanAcoleyen:2011mj}, and showed that the resulting actions are dilaton effective actions.} in Section \ref{sec:3}, where we also check whether the black hole solutions obtained from the naive limit still survives. Additionally, a comparison with an alternative method, ``the conformal trick'' of \cite{Mann:1992ar} is provided. We end our paper with conclusions in Section \ref{sec:4}.

\section{Naive Limit vs Limit via Regularized Kaluza-Klein Reduction}\label{sec:2}
In this section, we first present the naive limit of the static black hole solutions of third-order Lovelock gravity. Then, we show how a well-defined theory can be obtained by the KK procedure of \cite{Lu:2020iav,Kobayashi:2020wqy}.
\subsection{The Naive Limit of the Static Black Hole Solution}
The $D$-dimensional action corresponding to the $m$-th order Lovelock Lagrangian is
\begin{equation}
\label{eqn8}
S=\int \dif^Dx\, \sqrt{-g}\, c_m\, \mathcal{L}_m,
\end{equation}
where $\mathcal{L}_m$ gives a non-trivial contribution to field equations in $D\ge 2m+1$. When a $D$-dimensional metric ansatz is inserted into field equations, one obtains factors of $\epsilon=D-p$ with $p\le 2m$. By scaling the coupling constant as $c_m \rightarrow \dfrac{c_m}{\epsilon}$ and then taking the limit $\epsilon  \rightarrow 0$ yield a non-trivial contribution to $p$-dimensional field equations. As an example, we consider the static black hole ansatz
\begin{equation}
\label{eq:ansatz}
\dif s_D^2=-N(r)^2\left(k+\frac{r^2}{L^2}f(r)\right)\dif t^2+\frac{\dif r^2}{\left(k+\frac{r^2}{L^2}f(r)\right)}+r^2\dif l_{(D-2),k}^2,
\end{equation}
where $k=+1,0,-1$ correspond to spherical, planar and hyperbolic horizons as follows
\begin{equation}
\begin{split}
k=+1&: \qquad \dif l_{(D-2),k}^2=\dif\Omega^2_{(D-2)},\\
k=0&: \qquad \dif l_{(D-2),k}^2=\frac{1}{L^2}(\dif x_2^2+\cdots+\dif x_D^2),\\
k=-1&: \qquad \dif l_{(D-2),k}^2=\dif \Sigma^2_{(D-2)}.
\end{split}
\end{equation}
The action for the cubic Lovelock gravity is given by
\begin{equation}
\label{eqn10}
S=\int\dif^Dx\sqrt{-g}\left[\zeta \mathcal{L}_{m=0} +\sigma\mathcal{L}_{m=1}+\alpha L^2\mathcal{L}_{m=2}+\beta L^4\mathcal{L}_{m=3}\right],
\end{equation}
where we introduce a cosmological constant as the zeroth-order Lovelock Lagrangian
\begin{equation}
\label{eqn11}
\mathcal{L}_{m=0} =\frac{(D-1)(D-2)}{L^2},
\end{equation}
and choose the coupling constants as $c_m=(\zeta,\sigma,\alpha L^2, \beta L^4)$. Note that $\zeta=+1,-1$ correspond to negative and positive cosmological constants respectively. Inserting the ansatz \eqref{eq:ansatz} into the action \eqref{eqn10}, one obtains an effective action $S_\text{eff}=\int\dif rL_\text{eff}[N(r),f(r)]$. After finding the Euler-Lagrange equations for $N(r)$ and $f(r)$, setting $N(r)=1$ yields
\begin{equation}
\label{eqn12}
(D-2)\frac{d}{dr}\left[r^{D-1}(\z-\sigma f+\alpha(D-3)(D-4)f^2-\beta(D-3)(D-4)(D-5)(D-6)f^3)\right]=0.
\end{equation}
In $D\ge 7$, this gives an algebraic equation for $f(r)$
\begin{equation}
\label{eqn13}
\z-\sigma f+\alpha(D-3)(D-4)f^2-\beta(D-3)(D-4)(D-5)(D-6)f^3=\left(\frac{\omega}{r}\right)^{D-1},
\end{equation}
where $\omega$ is a constant. Depending on the coefficients, this cubic equation might have 1 real and 2 complex roots or 3 real roots. Therefore, it is even possible to obtain a unique solution unlike the EGB theory (see \cite{deBoer:2009gx,Camanho:2009hu,Myers:2010ru} for examples\footnote{This was shown to be the case for all odd-order Lovelock gravities in \cite{Camanho:2011rj}.}). One can easily see that important physical properties can be obtained directly from \eqref{eqn13}. For a planar horizon\footnote{For $k \neq 0$, see \cite{Myers:2010ru}.} ($k=0$) with $\z=1$, i.e., a negative cosmological constant in the action, the event horizon is located at $r=\omega$ and the temperature of the black hole is $T=\frac{\omega^2f^\pr(r=\omega)}{4 \pi L^2}=\frac{(D-1)\omega}{4 \pi \s L^2}$. Since the explicit form of the solution(s) is rather complicated, we will focus on the equation for the metric function $f(r)$ \eqref{eqn13} to check whether they survive within the theory obtained by our well-defined limit. 

Application of the naive limit to cubic Lovelock gravity gives lower-dimensional solutions in $p=2,3,4,5,6$. Scaling of the parameters, how we fix the coefficient of the Einstein-Hilbert term for simplicity, and the resulting equation for the metric function $f(r)$ are given in Table \ref{table:1}.

{\renewcommand{\arraystretch}{2.5} 
	\begin{table}[h]
		\centering
		\scalebox{0.9}{
			\begin{tabular} { |p{0.2cm}|p{8cm}|p{7cm}|  }
				\hline
				$p$ & Scaling and Fixing Parameters  & Equation for metric function $f(r)$     \\ \hline
				6   & $\beta\to\dfrac{1}{D-6}\ \beta\ ,\ \sigma=1$                                             & $\dfrac{\dif}{\dif r}\left[r^5(\zeta- f+6\alpha f^2-6\beta f^3)\right]=0$
				\\\hline
				5   & $\beta\to\dfrac{1}{D-5}\ \beta\ ,\ \sigma=1$                                             & $\dfrac{\dif}{\dif r}\left[r^4(\zeta- f+2\alpha f^2+2\beta f^3)\right]=0$ \\\hline
				4   & $(\alpha,\beta)\to\dfrac{1}{D-4}\ (\alpha,\beta)\ ,\ \sigma=1$                           & $\dfrac{\dif}{\dif r}\left[r^3(\zeta-f+\alpha f^2-2\beta f^3)\right]=0$  \\\hline
				3   & $(\alpha,\beta)\to\dfrac{1}{D-3}\ (\alpha,\beta)\ ,\ \sigma=1$                           & $\dfrac{\dif}{\dif r}\left[r^2(\zeta-f-\alpha f^2+6\beta f^3)\right]=0$  \\\hline
				2   & $(\zeta,\sigma,\alpha,\beta)\to\dfrac{1}{D-2}\ (\zeta,\sigma,\alpha,\beta)\ ,\ \sigma=1$ & $\dfrac{\dif}{\dif r}\left[r(\zeta-f+2\alpha f^2-24\beta f^3)\right]=0$  \\ \hline
		\end{tabular}}
		\caption{The naive limit of the $D$-dimensional equation \eqref{eqn12} for the metric function $f(r)$.}
		\label{table:1}
	\end{table}
	
	An interesting property of the naive limit is that it can also be applied to the trace of the field equations in certain dimensions as demonstrated for Einstein-Gauss-Bonnet theory in \cite{Glavan:2019inb}. For the $m$-th order Lovelock Lagrangian, the trace of the field equations takes a particularly simple form
	\begin{equation}
	\label{eqn14}
	\varepsilon^\mu_{(m)\mu}=\left(m-\frac{D}{2} \right)\mathcal{L}_m=0,
	\end{equation}
	which follows from \eqref{eq:1.2} by using the fact that $\sqrt{-g}\cL_{m}$ is a homogeneous function of the inverse metric $g^{\m\n}$ of degree $\left(m-\frac{D}{2}\right)$ \cite{Padmanabhan:2013xyr}. Applying to cubic Lovelock gravity gives $(\mbox{for} \,\zeta=1, \sigma=1)$,
	\begin{equation}
	\label{eqn15}
	D\mathcal{L}_{m=0}+(D-2)\mathcal{L}_{m=1}+\alpha L^2 (D-4)\mathcal{L}_{m=2}+\beta L^4(D-6) \mathcal{L}_{m=3}=0,
	\end{equation}
	which admits the naive limit $D\rightarrow 6$ after a scaling $\beta \rightarrow \frac{\beta}{D-6}$, yielding
	\begin{equation}
	\label{eqn16}
	\frac{120}{L^2}+4\mathcal{L}_{m=1}+2\alpha L^2 \mathcal{L}_{m=2}+\beta L^4 \mathcal{L}_{m=3}=0.
	\end{equation}
	When the solution from the naive limit survives in a tensor-scalar theory, this geometric constraint should also survive as a result of the trace of the gravitational field equation of the scalar equation. However, not all solutions of the constraint equation will satisfy the field equations. Therefore, it can only be used as a consistency check.
	
	\subsection{The Limit via Regularized Kaluza-Klein Reduction}\label{subsec:limit}
	In order to obtain a well-defined theory, one starts from the $D$-dimensional action \eqref{eqn10} and consider the following parametrization of the $D$-dimensional metric
	\begin{equation}
	\label{eqn17}
	\dif s_D^2=\dif s_p^2+e^{2\phi}\dif \Omega_{\epsilon,\lambda}^2,
	\end{equation}
	where the breathing scalar $\phi$ depends only on the external $p$-dimensional coordinates. $\dif s_p^2$ and $\dif \Omega_{\epsilon,\lambda}^2$ are the line elements of $p$-dimensional external spacetime and $\epsilon$-dimensional internal space respectively. The internal space is taken to be maximally symmetric, and therefore satisfies
	\begin{equation}
	\label{eqn18}
	R_{abcd}=\lambda(g_{ac} g_{bd}-g_{ad} g_{bc}),
	\end{equation}
	where the constant $\lambda$ characterizes the curvature of the internal space.
	The reduction of the $D$-dimensional action through this ansatz gives
	\begin{equation}
	\label{eqn19}
	S_p=c_m\int \dif^p x\, \sqrt{-g} e^{\epsilon\phi} \mathcal{L}_m (\epsilon),
	\end{equation}
	where $\mathcal{L}_m (\epsilon)$ is of the form $\mathcal{L}_m (\epsilon)=\mathcal{L}_m (\epsilon=0)+\Phi$. Here, $\mathcal{L}_m (\epsilon=0)$
	is just the $m$-th order Lovelock Lagrangian in $p$-dimensions, which identically vanishes or is a boundary term. The $\Phi$ term contains non-minimal coupling and pure scalar terms, which all carry an $\epsilon$ factor.
	Expanding the $p$-dimensional action in powers of $\epsilon$, we have
	\begin{equation}
	\label{eqn20}
	S_p=c_m\int \dif^px\sqrt{-g}\left[\mathcal{L}_m (\epsilon=0)+\frac{\dif}{\dif\ep}\big[e^{\ep \phi} \mathcal{L}_{m}(\ep)\big]\Bigr|_{\ep=0}\ep+\mathcal{O} (\ep^2)\right].
	\end{equation}
	After adding $-c_m\int\dif^p x\sqrt{-g}\, \mathcal{L}_m (\epsilon=0)$, which has no effect on the field equations, one can scale the coupling constant as $c_m \rightarrow \frac{c_m}{\epsilon}$ and take the limit $\ep
	\rightarrow 0$ ($D\goesto p$), which gives
	\begin{equation}
	\label{eqn21}
	S_p=c_m\int \dif^p x\sqrt{-g}\left[\frac{\dif}{\dif\ep}\big[e^{\ep \phi} \mathcal{L}_{m}(\ep)\big]\Bigr|_{\ep=0}\right].
	\end{equation}
	The resulting theory is described by a Lagrangian, which is just the $\mathcal{O}(\ep)$-term in the series expansion of $e^{\ep \phi} \mathcal{L}_m (\ep)$. In the next section, we will make use of this logic to obtain lower-dimensional limits of cubic Lovelock gravity and check whether the naive limit of the static black hole solution survives or not.

	\section{Lower-dimensional Limits of Cubic Lovelock Gravity}\label{sec:3}
	The procedure that we have described in subsection \ref{subsec:limit} was successfully applied for $m=1,2$ in \cite{Lu:2020iav,Kobayashi:2020wqy} to obtain the lower-dimensional limits of EGB theory, which have the following properties:
	\begin{itemize}
		\item As a result of the consistency of the KK procedure, the field equations remain second-order and one obtains scalar-tensor theories of Horndeski class.
		\item When the internal space is non-flat ($\lambda \neq 0)$, one obtains terms breaking the shift symmetry of the scalar field.
		\item For static field configurations, the curvature of the internal space does not affect the metric profile but changes the profile of the scalar field. The naive limit of the static black hole solution is preserved as $D \goesto p=2,3,4$.
	\end{itemize}
	Applying the KK procedure to the cubic Lovelock Lagrangian $\cL_{m=3}$ is considerably more cumbersome compared to $m=1,2$ cases. Therefore, we give the details and the most general form of the resulting $p$-dimensional Lagrangian in Appendix \ref{app:KK}, where we also present the results for $m=1,2$ Lagrangians for completeness \cite{Lu:2020iav,Kobayashi:2020wqy}.
	
	In this section, we will take a flat internal space ($\lambda=0$) for simplicity and show that the solution suggested by the naive limit is preserved with the exception that in 4D, only a planar horizon is allowed.
	
	\subsection{$p=6$}
	In $p=6$, $\cL_{m=3}$ is a boundary term. Therefore, the action for the 6D cubic Lovelock gravity is given by
	\begin{equation}
	\label{eq:p6}
	S=\int\dif^6x\sqrt{-g}\left[\zeta\frac{20}{L^2}+\mathcal{L}_{m=1}+\alpha L^2\mathcal{L}_{m=2}+\beta L^4\frac{\dif}{\dif\ep}\left[e^{\ep\phi}\mathcal{L}_{m=3}(\ep)\right]\Bigr|_{\ep=0}\right],
	\end{equation}
	where the lower-dimensional limit of the cubic Lovelock Lagrangian is
	\begin{align}\label{eq:cubicred}
	\frac{\dif}{\dif\ep}\left[e^{\ep\phi}\mathcal{L}_{m=3}(\ep)\right]\Bigr|_{\ep=0}&=\phi \mathcal{L}_{m=3}-3\mathcal{L}_{m=2}(\partial\phi)^2-12 R(\Box\phi)^2-48R_{\mu}\, ^{\delta} R_{\nu \delta}\nabla^\mu \phi\nabla^\nu\phi\nn\\
	&+24R_{\mu\nu} R\nabla^\mu \phi\nabla^\nu \phi-48R^{\delta\mu}R_{\nu\delta\rho\mu}\nabla^\nu \phi\nabla^\rho \phi+24 R_\alpha\, ^{\delta\mu\nu} R_{\beta\delta\mu\nu}\nabla^\alpha\phi\nabla^\beta \phi\nn\\
	&+6R\big((\partial\phi)^2\big)^2+24R\nabla_\mu \nabla_\nu\phi\nabla^\mu \phi\nabla^\nu \phi+12 R(\nabla_\mu \nabla_\nu \phi)^2\nn\\
	&-96R_\mu\,^\delta\nabla^\mu \phi\nabla^\nu \phi\nabla_\delta \nabla_\nu \phi+48 R^{\mu\nu}\Box\phi\nabla_\mu \nabla_\nu \phi+48R_{\mu\nu}\Box\phi\nabla^\mu \phi\nabla^\nu \phi\nn\\
	&-48R_{\mu\nu}(\partial \phi)^2\nabla^\mu \phi\nabla^\nu \phi-48R^{\mu\nu}\nabla_\delta \nabla_\nu \phi\nabla^\delta\nabla_\mu \phi\nn\\
	&-48R_{\alpha\delta\beta\mu}\nabla^\alpha \phi\nabla^\beta \phi\nabla^\mu \nabla^\delta \phi-24R_{\alpha\delta\beta\mu}\nabla^\beta \nabla^\alpha \phi\nabla^\mu \nabla^\delta \phi\nn\\
	&-16(\Box \phi)^3+24(\partial \phi)^2(\Box\phi)^2+96\Box \phi \nabla_\mu\nabla_\nu \phi\nabla^\mu \phi\nabla^\nu \phi-24\big((\partial \phi)^2\big)^3\nn\\
	&-144(\partial\phi)^2\nabla_\mu\nabla_\nu \phi\nabla^\mu \phi\nabla^\nu \phi-96\nabla^\mu\phi\nabla^\nu \phi\nabla_\delta\nabla_\nu \phi\nabla^\delta\nabla_\mu \phi\nn\\
	&-32\nabla^\mu\nabla^\nu \phi\nabla_\delta\nabla_\mu \phi\nabla^\delta\nabla_\nu \phi+48\Box\phi (\nabla_\mu\nabla_\nu \phi)^2-24(\partial \phi)^2 (\nabla_\mu\nabla_\nu \phi)^2.
	\end{align}
	To study static field configurations, we use the ansatz \eqref{eq:ansatz} and $\phi=\phi(r)$ in the action \eqref{eq:p6}. The Euler-Lagrange equations corresponding to the effective action $S_\text{eff}=\int\dif r\ L_\text{eff}[N(r),f(r),\phi(r)]$, are at most second-order as guaranteed by the consistency of the KK procedure. For $N(r)=1$, the equation that follows from the variation  of the metric function $f(r)$ is
	\begin{equation}\label{eq:p6phi}
	\dfrac{\delta S_\text{eff}}{\delta f}\Bigr|_{N=1}=0\Rightarrow \left[kL^2(-2+r\phi')+rf(-1+r\phi')^2\right](\phi'^2+\phi'')=0,
	\end{equation}
	which has two solutions. When $k \neq 0$, $\phi(r)=\log(r)$ is not consistent with the remaining equations. For the solution
	\begin{equation}\label{eq:p6phisol}
	\phi(r)=\log(r)+\int_1^{r}\dif x\ \frac{L}{x}\frac{|k|}{\sqrt{L^2+kx^2f(x)}},
	\end{equation}
	$\dfrac{\delta S_\text{eff}}{\delta \phi}\Bigr|_{N=1}=0$ is automatically satisfied, and $\dfrac{\delta S_\text{eff}}{\delta N}\Bigr|_{N=1}=0$ reduces to the equation for the metric function $f(r)$ from the naive limit as given in Table \ref{table:1}.
	
	\subsection{$p=5$}
	In $p=5$, $\cL_{m=3}$ vanishes identically. The action for the 5D cubic Lovelock gravity is given by
	\begin{equation}
	S=\int\dif^5x\sqrt{-g}\left[\zeta\frac{12}{L^2}+\sigma\mathcal{L}_{m=1}+\alpha L^2\mathcal{L}_{m=2}+\beta L^4\frac{\dif}{\dif\ep}\left[e^{\ep\phi}\mathcal{L}_{m=3}(\ep)\right]\Bigr|_{\ep=0}\right],
	\end{equation}
	where the only difference is that the first term in the reduced Lagrangian \eqref{eq:cubicred} is zero in this case. For $N(r)=1$, the equation that follows from the variation  of the metric function $f(r)$ is identical to the $p=6$ case, which is shown in \eqref{eq:p6phi}  and the consistent solution \eqref{eq:p6phisol} gives rise to the naive limit of the equation for the metric function $f(r)$ given in Table \ref{table:1}.
	
	\subsection{$p=4$}
	In $p=4$, $\cL_{m=2}$ is a boundary term and $\cL_{m=3}$ vanishes identically. Therefore the action for the 4D cubic Lovelock gravity is given by
	\begin{equation}
	S=\int \dif^4x\sqrt{-g}\left[\zeta\frac{6}{L^2}+\mathcal{L}_{m=1}+\alpha L^2\frac{\dif}{\dif\ep}\left[e^{\ep\phi}\mathcal{L}_{m=2}(\ep)\right]\Bigr|_{\ep=0}+\beta L^4\frac{\dif}{\dif\ep}\left[e^{\ep\phi}\mathcal{L}_{m=3}(\ep)\right]\Bigr|_{\ep=0}\right],
	\end{equation}
	where the lower-dimensional limit of the Gauss-Bonnet term is
	\begin{equation}
	\label{eq:gbred}
	\frac{\dif}{\dif\ep}\left[e^{\ep\phi}\mathcal{L}_{m=2}(\ep)\right]\Bigr|_{\ep=0}=\phi\cL_{m=2}+4G^{\mu\nu}\der_\mu\phi\der_\nu\phi-4(\partial\phi)^2\Box\phi+2((\partial\phi)^2)^2,
	\end{equation}
	and the lower-dimensional limit of the cubic Lovelock Lagrangian is given in \eqref{eq:cubicred}. For $N(r)=1$, the equation that follows from the variation of the metric function $\dfrac{\delta S_\text{eff}}{\delta f}\Bigr|_{N=1}
	=0$ implies
	\begin{equation}\label{eq:p4phieq}
	\begin{split}
	0=\Big\{&18\beta r^4f^2\phi'^2(-1+r\phi')^2+k^2L^2\phi'\Big(2k\alpha r+(-k\alpha r^2+12L^2\beta)\phi'-36L^2\beta r\phi'^2+18L^2\beta r^2\phi'^3\Big)\\
	&-r^2f\Big(\alpha-2\alpha r\phi'+(\alpha r^2-30kL^2\beta)\phi'^2+72kL^2\beta r\phi'^3-36kL^2\beta r^2\phi'^4\Big)\Big\}(\phi'^2+\phi''),
	\end{split}
	\end{equation}
	which is far more complicated  for $k\neq0$ when compared to the $p=5,6$ cases, since there are now contributions from both $m=2$ and $m=3$ terms. The various possibilities for the solutions are as follows
	\begin{itemize}
		\item \underline{$k=0$} : Eq. \eqref{eq:p4phieq} takes a very simple form
		\begin{equation}
		(-\alpha+18\beta r^2f\phi'^2)(\phi'^2+\phi'')=0.
		\end{equation}
		$\phi(r)=\log(r)$ is a consistent solution and reproduces the naive limit given in Table \ref{table:1}. The second possibility is
		\begin{equation}
		\phi(r)=\sqrt{\frac{\alpha}{18\beta}}\int_1^r\frac{\dif x}{x\sqrt{f(x)}},
		\end{equation}
		which is also a consistent solution and yields the following equation for the metric function $f(r)$
		\begin{equation}
		\begin{split}
		0&=\alpha ^3+2916 \beta ^2-54 \beta  \left(\alpha ^2+18 \beta\right) rf'+432 \sqrt{2} \beta ^3 \left(\frac{\alpha }{\beta }\right)^{3/2} rf^{1/2} f'\\
		&-162 \beta  \left(\alpha ^2+18 \beta \right)f+864 \sqrt{2} \beta ^3 \left(\frac{\alpha }{\beta }\right)^{3/2} f^{3/2}.
		\end{split}
		\end{equation}
		For generic values $\alpha$ and $\beta$, there is no analytical solution to this equation. Obviously, the solution will differ from the one suggested by the naive limit, and it might be interesting to look for numerical solutions.
		
		\item \underline{$k\neq0$} : $\phi(r)=\log(r)$ is not a consistent solution in this case. For generic values of $\alpha$ and $\beta$, there is no analytical solution to \eqref{eq:p4phieq}. If the solution from the naive limit is to be preserved, it should follow from the solution of the second branch. One can try to achieve this numerically for certain choices of $\alpha$ and $\beta$.
		
		However, by a simple check one can show that the naive limit is not recovered with the cubic term. When the limit of only the cubic term is considered, the naive limit suggests  $f(r)=\dfrac{a}{r}$, where $a$ is a constant. Inserting this into the equation following from the variation of the action $S_\text{eff}$ with respect to the metric function $f(r)$, one can solve for the scalar field as
		\begin{equation}
		\phi(r)=\log(r)\pm\frac{2}{\sqrt{3}}\text{atanh}\left[\sqrt{1+\frac{kar}{L^2}}\ \right],
		\end{equation}
		which is not consistent with the remaining equations $\dfrac{\delta S_\text{eff}}{\delta N}\Bigr|_{N=1}=0$ and $\dfrac{\delta S_\text{eff}}{\delta \phi}\Bigr|_{N=1}=0$.
	\end{itemize}
	
	\subsection{$p=3$}\label{subsec:3d}
	In $p=3$, $\cL_{m=2}$ and $\cL_{m=3}$ Lagrangians vanish identically. The action for the 3D cubic Lovelock gravity is
	\begin{equation}
	S=\int\dif^3x\sqrt{-g}\left[\zeta\frac{2}{L^2}+\mathcal{L}_{m=1}+\alpha L^2\frac{\dif}{\dif\ep}\left[e^{\ep\phi}\mathcal{L}_{m=2}(\ep)\right]\Bigr|_{\ep=0}+\beta L^4\frac{\dif}{\dif\ep}\left[e^{\ep\phi}\mathcal{L}_{m=3}(\ep)\right]\Bigr|_{\ep=0}\right],
	\end{equation}
	where the lower-dimensional limits of the Gauss-Bonnet and cubic Lovelock Lagrangians are given in \eqref{eq:gbred} and \eqref{eq:cubicred}. Due to the following decomposition of the Riemann tensor in 3D
	\begin{equation}
	R_{\mu\nu\rho\sigma}=2(g_{\mu[\rho}R_{\sigma]\nu}-g_{\nu[\rho}R_{\sigma]\mu})+Rg_{\mu[\rho}g_{\sigma]\nu},
	\end{equation}
	$m=3$ term simplifies to
	\begin{equation}
	\begin{split}
	\frac{\dif}{\dif\ep}\left[e^{\ep\phi}\mathcal{L}_{m=3}(\ep)\right]\Bigr|_{\ep=0}&=-48G^{\mu\nu}(\partial\phi)^2\nabla_\mu \nabla_\nu \phi+6R\big((\partial\phi)^2\big)^2-48R_{\mu\nu}(\partial\phi)^2\nabla^\mu \phi\nabla^\nu\phi\\
	&-16(\Box\phi)^3+24(\partial\phi)^2(\Box\phi)^2+96\Box\phi \nabla_\mu\nabla_\nu\phi\nabla^\mu \phi\nabla^\nu \phi-24\big((\partial\phi)^2\big)^3\\
	&-144(\partial\phi)^2\nabla_\mu\nabla_\nu \phi\nabla^\mu\phi\nabla^\nu\phi-96\nabla^\mu \phi\nabla^\nu \phi\nabla_\delta \nabla_\nu \phi\nabla^\delta \nabla_\mu \phi\\
	&-32\nabla^\mu \nabla^\nu \phi\nabla_\delta \nabla_\mu \phi\nabla^\delta \nabla_\nu\phi+48\Box\phi (\nabla_\mu \nabla_\nu \phi)^2-24(\partial\phi)^2(\nabla_\mu \nabla_\nu \phi)^2.
	\end{split}
	\end{equation}
	
	Since the event horizon is one-dimensional, the most general form of the static black hole ansatz is
	\begin{equation}
	\dif s_D^2=-N(r)^2\frac{r^2}{L^2}f(r)\dif t^2+\frac{\dif r^2}{\frac{r^2}{L^2}f(r)}+r^2\dif\theta^2.
	\end{equation}
	We now have
	\begin{equation}
	\dfrac{\delta S_\text{eff}}{\delta f}\Bigr|_{N=1}=0\Rightarrow(-\alpha+18\beta r^2f\phi'^2)(\phi'^2+\phi'')=0,
	\end{equation}
	and $\phi(r)=\log(r)$ gives the naive limit of the equation for the metric function given in Table \ref{table:1}. On the other hand, the solution of the other branch
	\begin{equation}
	\phi(r)=\sqrt{\frac{\alpha}{18\beta}}\int_1^r\frac{\dif x}{x\sqrt{f(x)}},
	\end{equation}
	yields the following equation for the metric function $f(r)$
	\begin{equation}
	\dfrac{\delta S_\text{eff}}{\delta N}\Bigr|_{N=1}=0\Rightarrow-\alpha^3-972\beta^2+54\beta(\alpha^2+18\beta)f+27\beta r(\alpha^2+18\beta)f'=0.
	\end{equation}
	The solution of this equation reads
	\begin{equation}
	f(r)=\frac{\alpha^3+972\beta^2}{54\alpha^2\beta+972\beta^2}-\frac{ML^2}{r^2},
	\end{equation}
	which describes the BTZ black hole \cite{Banados:1992wn} whose line element is given by
	\begin{equation}
	\dif s^2=-\left(\frac{r^2}{\el^2}-M\right)\dif t^2+\dfrac{\dif r^2}{\left(\frac{r^2}{\el^2}-M\right)}+r^2\dif\theta^2,
	\end{equation}
	with the length scale $\el$ determined by the equation
	\begin{equation}
	\dfrac{\el^2}{L^2}=\dfrac{54\alpha^2\beta+972\beta^2}{\alpha^3+972\beta^2}.
	\end{equation}
	
	\subsection{$p=2$}
	In $p=2$,  $\cL_{m=1}$ is a boundary term and $\cL_{m=2,3}$ vanish identically. The action for the 2D cubic Lovelock gravity is
	\begin{equation}
	\begin{split}
	S=\int\dif^2x\sqrt{-g}\Big[&\zeta\frac{1}{L^2}
	+\frac{\dif}{\dif\ep}\left[e^{\ep\phi}\mathcal{L}_{m=1}(\ep)\right]\Bigr|_{\ep=0}\\
	&+\alpha L^2\frac{\dif}{\dif\ep}\left[e^{\ep\phi}\mathcal{L}_{m=2}(\ep)\right]\Bigr|_{\ep=0}+\beta L^4\frac{\dif}{\dif\ep}\left[e^{\ep\phi}\mathcal{L}_{m=3}(\ep)\right]\Bigr|_{\ep=0}\Big],
	\end{split}
	\end{equation}
	where the two-dimensional limits are simplified as
	\begin{equation}
	\begin{split}
	\frac{\dif}{\dif\ep}\left[e^{\ep\phi}\mathcal{L}_{m=1}(\ep)\right]\Bigr|_{\ep=0}&=\phi R-(\partial\phi)^2,\\
	\frac{\dif}{\dif\ep}\left[e^{\ep\phi}\mathcal{L}_{m=2}(\ep)\right]\Bigr|_{\ep=0}&=-4(\partial\phi)^2\Box\phi+2((\partial\phi)^2)^2,\\
	\frac{\dif}{\dif\ep}\left[e^{\ep\phi}\mathcal{L}_{m=3}(\ep)\right]\Bigr|_{\ep=0}
	&=18R((\partial\phi)^2)^2-16(\Box\phi)^3+24(\partial\phi)^2(\Box\phi)^2+96\der_\mu\der_\nu\phi\der^\mu\phi\der^\nu\phi\Box\phi\\
	&-24((\partial\phi)^2)^3-144\der_\mu\der_\nu\phi\der^\mu\phi\der^\nu\phi(\partial\phi)^2-96\der^\mu\phi\der^\nu\phi\der_\delta\der_\nu\phi\der^\delta\der_\mu\phi\\
	&-32\der^\mu\der^\nu\phi\der_\delta\der_\mu\phi\der^\delta\der_\nu\phi+48(\der_\mu\der_\nu\phi)^2\Box\phi-24(\der_\mu\der_\nu\phi)^2(\partial\phi)^2,
	\end{split}
	\end{equation}
	due to the following decomposition of the Riemann tensor in 2D
	\begin{equation}
	R_{\mu\nu\rho\sigma}=\frac{R}{2}(g_{\mu\rho}g_{\nu\sigma}-g_{\mu\sigma}g_{\nu\rho}).
	\end{equation}
	
	The most general form of the static black hole ansatz in 2D is 
	\begin{equation}
	\dif s_D^2=-N(r)^2\frac{r^2}{L^2}f(r)\dif t^2+\frac{\dif r^2}{\frac{r^2}{L^2}f(r)}.
	\end{equation}
	The equation for the scalar field is given by
	\begin{equation}
	\dfrac{\delta S_\text{eff}}{\delta f}\Bigr|_{N=1}=0\Rightarrow(1-4\alpha r^2f\phi'^2+72\beta r^4f^2\phi'^4)(\phi'^2+\phi'')=0.
	\end{equation}
	$\phi(r)=\log(r)$ gives the naive limit as described in Table \ref{table:1}. The solution of the scalar field for the other branch implies $f(r)=0$.
	
	\subsection{Comparison with the Conformal Trick}
	A well-defined theory in lower dimensions can also be obtained by the so-called ``conformal trick'', which was first used to obtain the $D\to2$ limit of general relativity by Mann and Ross in \cite{Mann:1992ar}. It can be easily generalized to Lovelock gravity as follows: Taking a conformally related metric in $D$-dimensions $\tilde{g}_{\mu\nu}=e^{\psi}g_{\mu\nu}$, $(\mu,\nu:0,1,\cdots,D-1)$, we consider the following action
	\begin{equation}\label{eq:conformal_lagrangian}
	S_D=c_m\left[\int\dif^Dx\sqrt{-\tilde{g}}e^{\omega\psi}\widetilde{\mathcal{L}}_m-\int\dif^Dx\sqrt{-g}\mathcal{L}_m\right],
	\end{equation}
	where $\widetilde{\mathcal{L}}_m$ is the $m$-th order Lovelock Lagrangian evaluated for the metric $\tilde{g}_{\mu\nu}$, whose general form reads
	\begin{equation}
	\widetilde{\mathcal{L}}_m=e^{-m\psi}\left[\mathcal{L}_m+\Psi\right].
	\end{equation}
	Here, $\Psi$ contains non-minimal coupling and pure scalar terms. The $e^{\omega\psi}$ term is a compensating factor which might be needed to obtain a result without an exponential term. With these at hand, the $D$-dimensional action \eqref{eq:conformal_lagrangian} takes the following generic form
	\begin{equation}
	S_D=c_m\int\dif^D x\sqrt{-g}\left[e^{\ep\psi}(\mathcal{L}_m+\Psi)-\mathcal{L}_m\right],
	\end{equation}
	where $\ep=D/2-m+\omega$. When all the terms in $\Psi$ carry an $\ep$ factor\footnote{This requires integration by parts and the use of Bianchi identities.}, after a scaling of the coupling constant $c_m\to\frac{c_m}{\ep}$ one can take the limit $\ep\to0$, which gives
	\begin{equation}
	S_p=c_m\int \dif^p x\sqrt{-g}\left[\dfrac{\dif}{\dif\ep}\left[e^{\ep\psi}(\mathcal{L}_m+\Psi)\right]\Bigr|_{\ep=0}\right].
	\end{equation}

	For $m=2$, one can perform the $D\to3,4$ limits by choosing $\omega=1/2,0$ respectively, which remarkably yields the same Lagrangian density \cite{Hennigar:2020fkv,Hennigar:2020lsl,Fernandes:2020nbq}. From the KK perspective, this is not surprising at all since one takes a limit $D\to p$ where $p\leq 2m$ and obtains a single Lagrangian density for all $p$. Therefore, one is tempted to think that as long as one can take a limit $\ep\to0$ via the conformal trick, the resulting theory should be valid for all $p$. Indeed, for $m=3$, a straightforward calculation shows that all terms in $\Psi$ have a factor of $(D-5)$ (without integration by parts), and the limit $\ep\to0$ ($D \goesto 5$) can be taken by choosing $\omega=1/2$. The resulting Lagrangian is identical to ours after the following redefinitions
	\begin{equation}
	\psi\to-2\phi, \qquad \beta\to-\beta.
	\end{equation}
	
	\section{Summary and Discussions}\label{sec:4}
	In this paper, we have obtained the lower-dimensional limits of cubic Lovelock gravity through the KK procedure of \cite{Lu:2020iav,Kobayashi:2020wqy} and discovered scalar-tensor theories with second-order field equations. In order to make a comparison with the naive limit of \cite{Glavan:2019inb} based on the scaling of the couplings, we studied the static black hole solutions and showed that the solution from the naive limit survives with only one exception: In 4D, one obtains the same solution only for the case of a planar horizon. When the horizon is spherical or hyperbolic, the equation for the metric function becomes much more complicated; however, it can be studied numerically to check if admits novel solutions.

	Note that the Lagrangian densities that follow from the regularized KK reduction are scalar-tensor theories with second-order field equations also in the critical dimension and beyond ($D \geq 2m$). However, it turns out that when the static field configurations are studied, they do not admit black hole solutions. Although they lose their physical significance in this context, these theories might still have some interesting features in higher dimensions ($D \geq 2m$), which deserve further study.
	
	There are several directions to pursue in probing the properties of the theories that we present here. In this work, we have focused on only the simplest type of static black hole solutions. More general solutions with $N(r) \neq 1$ in the ansatz \eqref{eq:ansatz} can be studied at least numerically (see \cite{Lu:2020iav} for an example in 4D EGB theory). A systematic study of vacua, black hole and cosmological solutions, the spectrum of linearized fluctuations and holographic properties can be performed as done for EGB theory in lower dimensions in \cite{Ma:2020ufk}. From the considerations of EGB theory 	\cite{Hennigar:2020fkv,Hennigar:2020drx}, rotating generalizations of the black hole solutions and
	interesting properties in black hole thermodynamics are also expected for these theories.
	
	Finally, we would like to emphasize that there is one more class of lower-dimensional limit of the third-order Lovelock Lagrangian $\cL_{m=3}$, giving rise to different scalar-tensor theories with broken shift-symmetry even when one takes a flat internal space. As can be seen in \eqref{eq:leps}, when the $D$-dimensional reduction ansatz \eqref{eqn17} is inserted, all the terms in $\cL_{m=3}(\ep)$ other than the $p$-dimensional Lovelock Lagrangian carry an $\ep-1$ factor, which allows a limit where $\ep \goesto 1$ ($D\goesto p+1$). Together with different possibilities for taking the limit of $m=1,2$ Lagrangians, this makes it possible to obtain various scalar-tensor theories with second-order field equations.
	
	\section*{Acknowledgements}
	We would like to acknowledge the xAct \cite{xAct} and xTras \cite{Nutma:2013zea} packages which were used in various tensorial computations. We thank Y. Nakayama for bringing their work \cite{Matsumoto:2022fln} to our attention. G. D. O. is supported by T\"{U}B\.{I}TAK Grant No 118C587.
	

	\newpage
	\appendix
	\section{Kaluza-Klein Reduction for $\cL_{m}$}\label{app:KK}
	In this appendix we provide the decomposition of the Lovelock Lagrangians $\cL_{m}$ with the parametrization given in \eqref{eqn17}. We encourage the interested reader to check the Mathematica calculations at \href{https://github.com/gsuer/arxiv-2203.01811}{https://github.com/gsuer/arxiv-2203.01811}.
	
	\subsection{$m=3$}
	Inserting the KK ansatz \eqref{eqn17} into the $D$-dimensional third-order Lovelock Lagrangian gives
	\setlength{\jot}{10pt}
	\begin{align}\label{eq:leps}
	&\mathcal{L}_{m=3}(\ep)
	=\nn\\
	& \mathcal{L}_{m=3}+\ep(\ep-1)(\ep-2)(\ep-3)(\ep-4)(\ep-5)e^{-6\phi}\lambda^3+3\ep(\ep-1)(\ep-2)(\ep-3)e^{-4\phi}\lambda^2R\nn\\
	&-6\ep(\ep-1)(\ep-2)(\ep-3)(\ep-4)e^{-4\phi}\lambda^2\Box\phi-3\ep(\ep-1)(\ep-2)(\ep-3)^2(\ep-4)e^{-4\phi}\lambda^2(\partial\phi)^2\nn\\
	&+3\ep(\ep-1)e^{-2\phi}\lambda\cL_{m=2}-12\ep(\ep-1)(\ep-2)e^{-2\phi}\lambda R\Box\phi-6\ep(\ep-1)^2(\ep-2)e^{-2\phi}\lambda R(\partial\phi)^2\nn\\
	&+24\ep(\ep-1)(\ep-2)e^{-2\phi}\lambda R^{\mu\nu}\der_\mu\der_\nu\phi+12\ep(\ep-1)(\ep-2)(\ep-3)e^{-2\phi}\lambda(\Box\phi)^2\nn\\
	&+12\ep(\ep-1)(\ep-2)(\ep-3)e^{-2\phi}\lambda(\partial\phi)^2\Box\phi+24\ep(\ep-1)(\ep-2)e^{-2\phi}\lambda R^{\mu\nu}\der_\mu\phi\der_\nu\phi\nn\\
	&-24\ep(\ep-1)(\ep-2)(\ep-3)e^{-2\phi}\lambda\der_\mu\der_\nu\phi\der^\mu\phi\der^\nu\phi-12\ep(\ep-1)(\ep-2)(\ep-3)e^{-2\phi}\lambda(\der_\mu\der_\nu\phi)^2\nn\\
	&+3\ep(\ep-1)\cL_{m=2}(\partial\phi)^2+12\ep(\ep-1)R(\Box\phi)^2+12\ep^2(\ep-1)(\partial\phi)^2\Box\phi+48\ep(\ep-1)R_\mu^{\ \delta}R_{\delta\nu}\der^\mu\phi\der^\nu\phi\nn\\
	&-24\ep(\ep-1)RR^{\mu\nu}\der_\mu\phi\der_\nu\phi+48\ep(\ep-1)R^{\rho\sigma}R_{\mu\rho\nu\sigma}\der^\mu\phi\der^\nu\phi-24R_\mu^{\ \alpha\beta\gamma}R_{\nu\alpha\beta\gamma}\der^\mu\phi\der^\nu\phi\nn\\
	&+48\ep(\ep-1)R_{\mu\rho\nu\sigma}\der^\mu\phi\der^\nu\phi\der^\rho\der^\sigma\phi+48\ep(\ep-1)R_{\mu\rho\nu\sigma}\der^\mu\der^\nu\phi\der^\rho\der^\sigma\phi\nn\\
	&+3\ep(\ep+1)(\ep-1)(\ep-2)R((\partial\phi)^2)^2-24\ep(\ep-1)R\der_\mu\der_\nu\phi\der^\mu\phi\der^\nu\phi-12\ep(\ep-1)R(\der_\mu\der_\nu\phi)^2\nn\\
	&-24\ep^2(\ep-1)R^{\mu\nu}\der_\mu\der_\nu\phi(\partial\phi)^2+96\ep(\ep-1)R_\alpha^{\ \delta}\der^\alpha\phi\der^\nu\phi\der_\delta\der_\nu\phi-48\ep(\ep-1)R^{\mu\nu}\der_\mu\der_\nu\phi\Box\phi\nn\\
	&-8\ep(\ep-1)(\ep-2)(\Box\phi)^3-12\ep(\ep-1)^2(\ep-2)(\partial\phi)^2(\Box\phi)^2-48\ep(\ep-1)R_{\mu\nu}\der^\mu\phi\der^\nu\phi\Box\phi\nn\\
	&-6\ep^2(\ep-1)(\ep-2)(\ep-3)((\partial\phi)^2)^2\Box\phi+48\ep(\ep-1)(\ep-2)\der_\mu\der_\nu\phi\der^\mu\phi\der^\nu\phi(\partial\phi)^2\nn\\
	&-24\ep(\ep-1)(\ep-2)R_{\mu\nu}\der^\mu\phi\der^\nu\phi(\partial\phi)^2-\ep(\ep+1)(\ep-1)(\ep-2)(\ep-3)(\ep-4)((\partial\phi)^2)^3\nn\\
	&+48\ep(\ep-1)R^{\mu\nu}\der_\delta\der_\mu\phi\der^\delta\der_\nu\phi+24\ep(\ep-1)(\ep-2)(\ep-3)\der_\mu\der_\nu\phi\der^\mu\phi\der^\nu\phi(\partial\phi)^2\nn\\
	&-48\ep(\ep-1)(\ep-2)\der^\mu\phi\der^\nu\phi\der_\delta\der_\mu\phi\der^\delta\der_\nu\phi-16\ep(\ep-1)(\ep-2)\der^\mu\der^\nu\phi\der_\delta\der_\mu\phi\der^\delta\der_\nu\phi\nn\\
	&+24\ep(\ep-1)(\ep-2)(\der_\mu\der_\nu\phi)^2\Box\phi+12\ep(\ep-1)^2(\ep-2)(\der_\mu\der_\nu\phi)^2(\partial\phi)^2,
	\end{align}
	where we have used integration by parts and Bianchi identities in the action \eqref{eqn19} to achieve this form.
	
	The $\ep\to0$ limit of the third-order Lovelock Lagrangian is given by
	\begin{align}
	&\frac{\dif}{\dif\ep}\left[e^{\ep\phi}\mathcal{L}_{m=3}(\ep)\right]\Bigr|_{\ep=0}=\nn\\
	&\phi \mathcal{L}_{m=3}-120e^{-6\phi}\lambda^3-18e^{-4\phi}\lambda^2 R-144e^{-4\phi}\lambda^2\Box\phi+216e^{-4\phi}\lambda^2(\partial\phi)^2\nn\\
	&-3e^{-2\phi}\lambda\cL_{m=2}-24e^{-2\phi}\lambda R\Box\phi+12e^{-2\phi} \lambda R(\partial\phi)^2+48e^{-2\phi} \lambda R^{\mu\nu}\nabla_\mu\nabla_\nu \phi\nn\\
	&-72e^{-2\phi} \lambda(\Box \phi)^2+144e^{-2\phi} \lambda(\partial\phi)^2\Box \phi+48e^{-2\phi} \lambda R_{\mu\nu} \nabla^\mu\phi\nabla^\nu\phi\nn\\
	&-72e^{-2\phi} \lambda\big((\partial\phi)^2\big)^2+144e^{-2\phi} \lambda\nabla_\mu \nabla_\nu \phi\nabla^\mu \phi\nabla^\nu \phi+72e^{-2\phi} \lambda(\nabla_\mu \nabla_\nu \phi)^2
	\nn\\
	&-3\cL_{m=2}(\partial\phi)^2-12 R(\Box\phi)^2-48R_{\mu}\, ^{\delta} R_{\nu \delta}\nabla^\mu \phi\nabla^\nu \phi+24R_{\mu\nu} R\nabla^\mu \phi\nabla^\nu \phi\nn\\
	&-48R^{\delta\mu}R_{\nu\delta\rho\mu}\nabla^\nu \phi\nabla^\rho \phi+24 R_\alpha\, ^{\delta\mu\nu} R_{\beta\delta\mu\nu}\nabla^\alpha \phi\nabla^\beta \phi+6R\big((\partial\phi)^2\big)^2\nn\\
	&+24R\nabla_\mu \nabla_\nu\phi\nabla^\mu \phi\nabla^\nu \phi+12R(\nabla_\mu \nabla_\nu \phi)^2-96R_\mu\,^\delta \nabla^\mu \phi\nabla^\nu \phi\nabla_\delta \nabla_\nu \phi\nn\\
	&+48 R^{\mu\nu}\Box\phi\nabla_\mu \nabla_\nu \phi+48R_{\mu\nu}\Box\phi\nabla^\mu \phi\nabla^\nu \phi-48R_{\mu\nu}(\partial \phi)^2\nabla^\mu \phi\nabla^\nu \phi\nn\\
	&-48R^{\mu\nu}\nabla_\delta \nabla_\nu \phi\nabla^\delta\nabla_\mu \phi-48R_{\alpha\delta\beta\mu} \nabla^\alpha \phi\nabla^\beta \phi\nabla^\mu \nabla^\delta \phi\nn\\
	&-24R_{\alpha\delta\beta\mu}\nabla^\beta\nabla^\alpha \phi\nabla^\mu\nabla^\delta\phi-16(\Box \phi)^3+24(\partial \phi)^2(\Box\phi)^2-24\big((\partial \phi)^2\big)^3\nn\\
	&+96\Box \phi\nabla^\mu \phi\nabla_\mu\nabla_\nu \phi\nabla^\nu\phi-32\nabla^\mu\nabla^\nu \phi\nabla_\delta\nabla_\mu \phi\nabla^\delta\nabla_\nu \phi\nn\\
	&-144(\partial\phi)^2\nabla_\mu\nabla_\nu \phi\nabla^\mu \phi\nabla^\nu \phi-96\nabla^\mu \phi\nabla^\nu \phi\nabla_\delta\nabla_\nu \phi\nabla^\delta\nabla_\mu \phi\nn\\
	&+48\Box\phi (\nabla_\mu\nabla_\nu \phi)^2-24(\partial \phi)^2 (\nabla_\mu\nabla_\nu \phi)^2.
	\end{align}
	
	For $p=3$ one has
	\begin{align}
	&\frac{\dif}{\dif\ep}\left[e^{\ep\phi}\mathcal{L}_{m=3}(\ep)\right]\Bigr|_{\ep=0}=\nn\\
	&-120e^{-6\phi}\lambda^3-18e^{-4\phi}\lambda^2 R-144e^{-4\phi}\lambda^2\Box\phi+216e^{-4\phi}\lambda^2(\partial\phi)^2\nn\\
	&-24e^{-2\phi}\lambda R\Box\phi+12e^{-2\phi}\lambda R(\partial\phi)^2+48e^{-2\phi} \lambda R^{\mu\nu}\nabla_\mu\nabla_\nu\phi-72e^{-2\phi} \lambda(\Box \phi)^2\nn\\
	&+144e^{-2\phi} \lambda(\partial\phi)^2\Box \phi+48e^{-2\phi} \lambda R_{\mu\nu}\nabla^\mu \phi\nabla^\nu \phi-72e^{-2\phi} \lambda\big((\partial\phi)^2\big)^2\nn\\
	&+144e^{-2\phi} \lambda
	\nabla_\mu \nabla_\nu \phi\nabla^\mu \phi\nabla^\nu \phi-72e^{-2\phi} \lambda (\nabla_\mu \nabla_\nu \phi) ^2+24 R (\partial\phi)^2\Box \phi\nn\\
	&+6R\big((\partial\phi)^2\big)^2-48R^{\mu\nu}(\partial\phi)^2\nabla_\mu \nabla_\nu \phi-48R_{\mu\nu}(\partial\phi)^2\nabla^\mu \phi\nabla^\nu \phi-16(\Box\phi)^3\nn\\
	&+24(\partial\phi)^2(\Box\phi)^2+96\Box\phi\nabla_\mu \nabla_\nu \phi\nabla^\mu \phi\nabla^\nu \phi-24\big((\partial\phi)^2\big)^3\nn\\
	&-144(\partial\phi)^2\nabla_\mu \nabla_\nu \phi\nabla^\mu \phi\nabla^\nu\phi-96\nabla^\mu \phi\nabla^\nu \phi\nabla_\delta \nabla_\nu \phi\nabla^\delta \nabla_\mu \phi\nn\\
	&-32\nabla^\mu \nabla^\nu\phi\nabla_\delta \nabla_\mu \phi\nabla^\delta\nabla_\nu\phi+48\Box\phi (\nabla_\mu \nabla_\nu \phi)^2-24(\partial\phi)^2(\nabla_\mu \nabla_\nu \phi)^2,
	\end{align}
	after using the decomposition of the Riemann tensor in three dimensions
	\begin{equation}
	R_{\mu\nu\rho\sigma}=2(g_{\mu[\rho}R_{\sigma]\nu}-g_{\nu[\rho}R_{\sigma]\mu})+Rg_{\mu[\rho}g_{\sigma]\nu}.
	\end{equation}
	
	For $p=2$, the limit reduces to the Lagrangian
	\begin{align}
	&\frac{\dif}{\dif\ep}\left[e^{\ep\phi}\mathcal{L}_{m=3}(\ep)\right]\Bigr|_{\ep=0}=\nn\\
	&-120e^{-6\phi}\lambda^3-18e^{-4\phi}\lambda^2 R-144e^{-4\phi}\lambda^2\Box\phi+216e^{-4\phi}\lambda^2(\partial\phi)^2\nn\\
	&+36e^{-2\phi}\lambda R(\partial\phi)^2-72e^{-2\phi}\lambda(\Box\phi)^2+144e^{-2\phi}\lambda(\partial\phi)^2\Box\phi-72e^{-2\phi}\lambda((\partial\phi)^2)^2\nn\\
	&+144e^{-2\phi}\lambda\der_\mu\der_\nu\phi\der^\mu\phi\der^\nu\phi+72e^{-2\phi}\lambda(\der_\mu\der_\nu\phi)^2+18R((\partial\phi)^2)^2-16(\Box\phi)^3\nn\\
	&+24(\partial\phi)^2(\Box\phi)^2+96\der_\mu\der_\nu\phi\der^\mu\phi\der^\nu\phi\Box\phi-24((\partial\phi)^2)^3\nn\\
	&-144\der_\mu\der_\nu\phi\der^\mu\phi\der^\nu\phi(\partial\phi)^2-96\der^\mu\phi\der^\nu\phi\der_\delta\der_\nu\phi\der^\delta\der_\mu\phi\nn\\
	&-32\der^\mu\der^\nu\phi\der_\delta\der_\nu\phi\der^\delta\der_\mu\phi+48(\der_\mu\der_\nu\phi)^2\Box\phi-24(\der_\mu\der_\nu\phi)^2(\partial\phi)^2,
	\end{align}
	due to the decomposition of the Riemann tensor in $2$-dimensions
	\begin{equation}
	R_{\mu\nu\rho\sigma}=\frac{R}{2}(g_{\mu\rho}g_{\nu\sigma}-g_{\mu\sigma}g_{\nu\rho}).
	\end{equation}
	
	\subsection{$m=2$}
	Inserting the KK ansatz \eqref{eqn17} into the $D$-dimensional second-order Lovelock Lagrangian gives
	\begin{align}
	\mathcal{L}_{m=2}(\ep)
	&=\mathcal{L}_{m=2}+\ep(\ep-1)(\ep-2)(\ep-3)e^{-4\phi}\lambda^2+2\ep(\ep-1)(\ep-2)(\ep-3)e^{-2\phi}\lambda (\partial\phi)^2\nn\\
	&+2\ep(\ep-1)e^{-2\phi}\lambda R-4\ep(\ep-1)G^{\mu\nu}\der_\mu\phi\der_\nu\phi-2\ep(\ep-1)(\ep-2)(\partial\phi)^2\Box\phi\nn\\
	&-\ep(\ep-1)^2(\ep-2)((\partial\phi)^2)^2.
	\end{align}
	
	The $\ep\to0$ limit of the second-order Lovelock Lagrangian is given by
	\begin{align}
	\frac{\dif}{\dif\ep}\left[e^{\ep\phi}\mathcal{L}_{m=2}(\ep)\right]\Bigr|_{\ep=0}&=\phi\cL_{m=2}-6e^{-4\phi}\lambda^2-2e^{-2\phi}\lambda R-12e^{-2\phi}\lambda(\partial\phi)^2\nn\\
	&+4G^{\mu\nu}\der_\mu\phi\der_\nu\phi-4(\partial\phi)^2\Box\phi+2((\partial\phi)^2)^2.
	\end{align}
	
	For $p=2$, it simplifies to
	\begin{equation}
	\frac{\dif}{\dif\ep}\left[e^{\ep\phi}\mathcal{L}_{m=2}(\ep)\right]\Bigr|_{\ep=0}=-6e^{-4\phi}\lambda^2-2e^{-2\phi}\lambda R-12e^{-2\phi}\lambda(\partial\phi)^2-4(\partial\phi)^2\Box\phi+2((\partial\phi)^2)^2.
	\end{equation}
	
	\subsection{$m=1$}
	Inserting the KK ansatz \eqref{eqn17} into the $D$-dimensional first-order Lovelock Lagrangian gives
	\begin{equation}
	\mathcal{L}_{m=1}(\ep)=\mathcal{L}_{m=1}+\ep(\ep-1)\lambda e^{-2\phi}+\ep(\ep-1)(\partial\phi)^2.
	\end{equation}
	The $\ep\to0$ limit of the second-order Lovelock Lagrangian is given by
	\begin{equation}
	\frac{\dif}{\dif\ep}\left[e^{\ep\phi}\mathcal{L}_{m=1}(\ep)\right]\Bigr|_{\ep=0}=\cL_{m=1}\phi-e^{-2\phi}\lambda-(\partial\phi)^2.
	\end{equation}
	\singlespacing

\end{document}